**Title:** Latent variation in pathogen strain-specific effects under multiple-versions-of-treatment theory

**Authors:** Bronner P. Gonçalves[1]

**Affiliations:**

[1] Faculty of Health and Medical Sciences, University of Surrey, Guildford, United Kingdom

**Correspondence:**

Faculty of Health and Medical Sciences, University of Surrey, 30 Priestley Rd, Guildford GU2 7YH, United Kingdom, bronnergoncalves@gmail.com

**Abstract**

Evidence-informed policy on infections requires estimates of their effects on health. However, pathogenic variation, whereby occurrence of adverse outcomes depends on the infecting strain, might complicate the study of many infectious agents. Here, we consider the interpretation of epidemiologic studies on effects of infections on health when there is heterogeneity in strain-specific effects and information on strain composition is unavailable. We use potential outcomes and causal inference theory for analyses in the presence of multiple versions of treatment to argue that oft-reported quantities in these studies have a causal interpretation that depends on population frequencies of infecting strains. Moreover, as in other contexts where the treatment-variation-irrelevance assumption might be violated, transportability requires additional considerations, beyond those needed for non-compound exposures. This discussion, that considers potential heterogeneity in strain-specific effects, will facilitate interpretation of these studies, and for the reasons mentioned above, also highlights the value of pathogen subtype data.

**Keywords:** causal inference, consistency, potential outcomes, infectious diseases

**Latent variation in pathogen strain-specific effects under multiple-versions-of-treatment theory**

*Background*

For many pathogens, effects of infections on health might vary depending on the infecting strain. Examples of such variation in strain-specific effects are numerous in epidemiology: evidence suggests heterogeneity in the risk of postacute sequelae after infections by different SARS-CoV-2 variants (1); colonisation by some *Group B Streptococcus* serotypes has been linked to higher risk of invasive infections (2); severe manifestations of malaria might be associated with virulent parasite strains (3).

Here, we consider the possibly frequent situation where variation in strain-specific effects is present but not directly observed; that is, the situation where data on pathogen presence are available, but in which strain information is not (e.g., because the detection assay does not discriminate between strains). In discussing this type of study, we use causal concepts with the objective of presenting a coherent interpretation of this scenario, with latent variation in strain-specific effects. In particular, we consider ideas developed for causal analyses with multiple versions of treatment (4-6), which, we argue, are useful to understand such studies, with compound exposures, that is, exposures for which version variation irrelevance (relative to a particular outcome) does not hold (when exposure variation irrelevance holds, the exposure is non-compound).

Before proceeding, we make two clarifications. First, although several terms, with different meanings, describe pathogen subtypes (variant, serotype, genotype, clone), throughout we, somewhat loosely, refer to these using the term strain (see introduction in (7) for discussion on possible categorizations of pathogen variation). Note also that the question considered here is distinct from that on potentially different vaccine effects on different strains (7-9).

*Notation*

Most causal inference analyses employ the consistency condition (see chapter 3 in (10) for a detailed discussion), which links the observed outcome of an individual to the potential outcome under the exposure level that she or he actually received. Now, if exposure corresponds to infection by a particular pathogen and if some pathogen strains are more virulent than others, then not only would there be multiple versions of the exposure, but also, as virulence varies, exposure variation irrelevance would not hold. Put differently, if we conceive of infections by different strains as constituting different ways of becoming infected with the pathogen, then, in the presence of variable pathogenicity, the consistency condition would not apply.

To discuss this issue, we consider an observational study where $A$, exposure, denotes infection by a pathogen detected by a method that does not discriminate between strains (1 = presence and 0 = absence of infection), and $Y$ denotes the outcome (e.g., 1 = hospitalization and 0 = no hospitalization during follow-up). Further, let $K$ denote version of pathogen exposure: either absence of infection, $K = 0$, or the infecting strain, in which case $K$ takes values from a set $\Phi$ of possible strains. Thus, $A$ can be considered a coarsened version of $K$ (4). We also define $Y^k$ as the potential outcome (11) of $Y$ had $K$ been set to $k$.

Interpretation of estimands in contexts with exposure variation that impacts potential outcomes requires additional considerations, and was discussed in detail by VanderWeele and Hernán (4). In particular, in that study, exposure variation irrelevance is defined, using potential outcomes that imply interventions on $A$ and $K$, as $Y^{a,k} = Y^{a,k'}$ for all individuals, all $a$ and all $k$ and $k'$ that represent possible versions of exposure level $a$. Here, this is assumed not to hold and we apply the framework in (4) to this scenario.

It is worth noting that if co-infections are possible, these could be viewed as additional exposure versions; $\Phi$ would then include infections with each strain, two-strain co-infections, three-strain co-infections, and so on. Finally, note that while we focus on versions of infection defined by strain information, relevant exposure versions might be linked to other aspects of infection such as duration; for instance, in a simplistic scenario, there could be short (relative a pre-specified duration) and long infections. In practice, infection duration information might be difficult to collect, as it requires frequent follow-up.

*Possible causal structure and interpretation of an oft-reported empirical quantity*

Panel A of **Figure 1** shows a causal diagram representing possible relationships between variables defined above. As shown in the figure, we assume that version ($K$) precedes exposure ($A$) (6), and conditional on $K$, $A$ and $Y$ are d-separated; this requires, for example, that all strains affecting $Y$ are included in $\Phi$. Notice that the diagram includes a common cause of $K$ and $Y$, denoted by $L$ (e.g., age could affect pathogen exposure and outcome); as mentioned above, $A$ could be viewed as a coarsened version of $K$, and for this reason, we present only the version-outcome confounder $L$ (see also Figure 5 in (6)). Note however that if the relation between $A$ and $K$ were non-deterministic, there could exist treatment-version confounders, which would need to be represented in the diagram. Further, since **Figure 1** implies that only $K$ affects $Y$, potential outcomes below are defined relative to $K$, $Y^k$.

Given the exchangeability assumption $Y^k \perp\!\!\!\perp K|L$, in this setting with multiple versions of infection, the often-reported quantity $\sum_l E[Y|A=1, L=l] \Pr(L=l) - \sum_l E[Y|A=0, L=l] \Pr(L=l)$, which does not require information on $K$, can be interpreted (4, 12) as a comparison that would be estimated in a trial where, within strata defined by $L$, for each arm, exposure version (that is, strain type or absence of infection) is randomly assigned based on relative frequencies of the versions by levels of $A$ and $L$ (see Proposition 8 in (4)):

$$\sum_l E[Y|A=1, L=l] \Pr(L=l) - \sum_l E[Y|A=0, L=l] \Pr(L=l)$$

$$= \sum_{l,k} E[Y^k|L=l] \Pr(K=k|A=1, L=l) \Pr(L=l)$$

$$- \sum_{l,k} E[Y^k|L=l] \Pr(K=k|A=0, L=l) \Pr(L=l)$$

If we assume that there is only one version of absence of infection, the expression could be written as:

$$\sum_l E[Y|A=1, L=l] \Pr(L=l) - \sum_l E[Y|A=0, L=l] \Pr(L=l)$$

$$= \sum_{l,k} E[Y^k|L=l] \Pr(K=k|A=1, L=l) \Pr(L=l) - \sum_l E[Y^0|L=l] \Pr(L=l)$$

$$(\because \Pr(K=0|A=0, L=l) = 1)$$

$$= \sum_{l,k} \{E[Y^k|L=l] - E[Y^0|L=l]\} \Pr(K=k|A=1, L=l) \Pr(L=l)$$

Thus, within levels of $L$, the expression corresponds to an average of strain-specific effects weighted by frequencies of the strains in strata defined $A=1$ and $L=l$. For illustration, if $Y$ denotes hospitalisation during the study, $A$ denotes SARS-CoV-2 infection detected by a test that does not provide information on variants ($K$), which might have different effects on $Y$ (13), and $L$ denotes age, then, if $L$ suffices to control for version-outcome confounding, the empirical quantity above could be viewed as an effect estimate from a randomized study where, conditional on $L$, participants in the arm assigned to infection are assigned to a particular variant with probability determined by the population-level distribution of variants among those infected; this group would be compared to the arm corresponding to absence of infection.

In Panels B and C of **Figure 1**, we present two hypothetical scenarios for a pathogen with three circulating strains in the population. While infections by strain 1 and strain 3 have effects, of different magnitudes, on $Y$, strain 2 does not to affect $Y$. In these panels, strain composition among infected participants is also shown. As indicated, despite assuming similar strain-specific effects in the two scenarios, $\sum_l E[Y|A=1, L=l] \Pr(L=l) - \sum_l E[Y|A=0, L=l] \Pr(L=l)$ has a higher numerical value when the relative frequency of strain 2 is lower.

*Discussion*

The expressions above suggest an interpretation of studies with latent heterogeneity in strain-specific effects that is coherent with causal inference theory. In particular, when we assume a single version of absence of infection, then $\sum_l E[Y|A=1, L=l] \Pr(L=l) - \sum_l E[Y|A=0, L=l] \Pr(L=l)$ could be viewed as an average of $l$-level strain-specific effects (relative to absence of infection) weighted by the distribution of $L$ and strain composition among those infected in $l$-defined strata. Moreover, analogous to (6), where authors argued that applying effects of a compound exposure estimated in one population to another is complicated by its dependence on exposure version distribution, even when strain-specific effects,

$E[Y^k - Y^0 | L = l]$, are transportable between two populations, $\sum_l E[Y|A = 1, L = l] \Pr(L = l) - \sum_l E[Y|A = 0, L = l] \Pr(L = l)$ might be different, unless distributions of versions $k$, within levels of $L$, are similar in the populations. Thus, when the goal is to transport average infection effects to different populations, or estimate strain-specific effects, strain composition data are needed. Of note, had there not been multiple versions of infection that could lead to different outcomes, $A$ would be a non-compound exposure, and, under consistency and conditional exchangeability, $\sum_l E[Y|A = 1, L = l] \Pr(L = l) - \sum_l E[Y|A = 0, L = l] \Pr(L = l)$ would have the interpretation of the effect of $A$ on $Y$.

Another implication of our discussion is that with temporal changes in strain composition (e.g., during periods following emergence of SARS-CoV-2 variants), the often-reported quantity discussed above might vary over time, regardless of changes in population-level immunity, as $\Pr(K = k | A = 1, L = l)$ would change for some $k$.

Related to arguments in (4), note that violation of the no-multiple-versions-of-treatment assumption is dependent on the outcome under consideration. For instance, it could be the case that for the outcome "fever during follow-up", potential outcomes under infection would not depend on the strain; however, if the outcome was hospitalization, then pathogen exposure variation might be relevant to potential outcomes, and the discussion above would apply.

Overall, the causal inference approach for studies with multiple versions of treatment provides a coherent interpretation of analyses with latent heterogeneity of strain-specific effects. In particular, this approach shows that in the presence of multiple treatment versions, it is possible to obtain the average effect in the study population under a slightly different interpretation which acknowledges the population-level distribution of versions but does not require strain data.

**Figure 1.** Epidemiologic setting with multiple versions of pathogen exposure. Panel A, which is similar to Figure 1 in (12), shows a causal diagram where $K$ indicates the infecting strain (or, for $K = 0$, absence of all pathogen strains), $A$ denotes overall infection status, $L$, a confounder, and $Y$, the outcome of interest. Panels B and C illustrate the impact of having different population level distributions of strains (that is, $\Pr(K = k|A = 1)$) on the quantity $E[Y|A = 1] - E[Y|A = 0]$ (for Panels B and C, we assume that there are no confounders, and hence the terms are not stratified). In these panels, we assume the presence of three strains, represented by $S_1, S_2,$ and $S_3$. Finally, note that the causal structure in Figure 4B in (12), which was presented in the context of psychosocial constructs, has similarities to the setting discussed here.

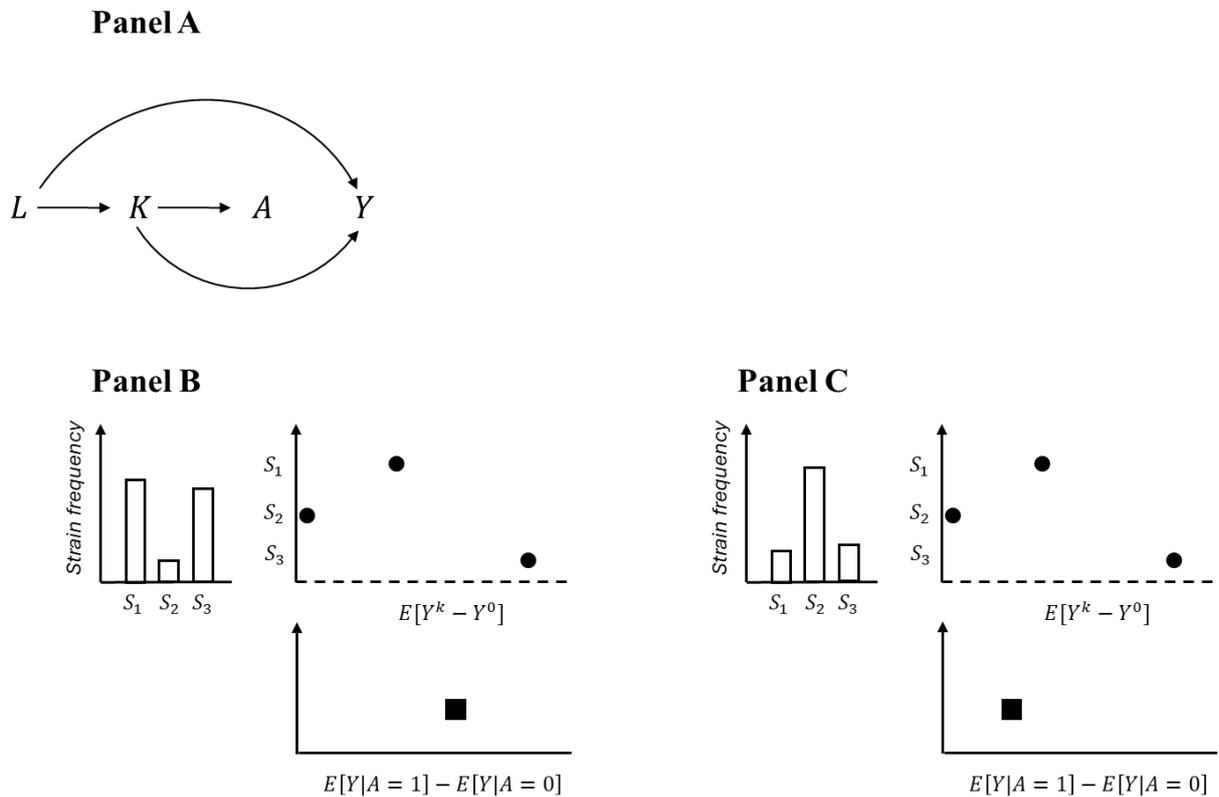